\documentclass[aps,twocolumn,pre,showpacs]{revtex4}

\usepackage{epsfig}

\newcommand{\e}  { {\rm e}}
\newcommand{\lb}  { l_{\rm B}}
\newcommand{\lp}  { l_{\rm p}}
\newcommand{\dd}  { {\rm d}}
\newcommand{\ele} { l_{\rm e}}
\newcommand{\kb}  { k_{\rm B}}
\newcommand{\br} { {\bf r} }
\newcommand{\bR} { {\bf R} }
\newcommand{\nM}  { n^{\rm M}}
\newcommand{\vdh} { V^{\rm DH} }
\newcommand{\debh} { {\rm DH} }
\newcommand{\phib} { \phi^{\rm b} }
\newcommand{\psib} { \psi^{\rm b} }
\newcommand{\phim} { \phi^{\rm m} }
\newcommand{\psim} { \psi^{\rm m} }
\newcommand{\nb} { n_0 }
\newcommand{\bk} { {\bf k} }

\begin{document}


\title{The Persistence Length of a Strongly Charged
Rod-like Polyelectrolyte \\ in Presence of Salt}
\author{Gil Ariel and  David Andelman}
\email{andelman@post.tau.ac.il}
\affiliation{
School of Physics and Astronomy, \\
Raymond and Beverly Sackler Faculty of Exact Sciences \\
Tel Aviv University, Tel Aviv 69978, Israel}

\date{October 20, 2002}

\begin{abstract}
\noindent
The persistence length of a single, intrinsically rigid
polyelectrolyte chain, above the Manning condensation threshold is
investigated theoretically in presence of added salt.
Using a loop expansion method, the partition function
is consistently calculated, taking into account corrections to mean-field theory.
Within a mean-field approximation,
the well-known results of Odijk, Skolnick and Fixman are reproduced.
Beyond mean-field, it is found that density
correlations between counterions and thermal fluctuations reduce
the stiffness of the chain, indicating an effective attraction
between monomers for highly charged chains and multivalent counterions.
This attraction results in a possible
mechanical instability (collapse), alluding to the phenomenon of DNA condensation.
In addition, we find that more counterions condense
on slightly bent conformations of the chain than predicted by the Manning model
for the case of an infinite cylinder.
Finally, our results are compared with previous models and experiments.

\end{abstract}


\maketitle


\noindent


\section{Introduction}
\label{intro}
\setcounter{equation}{0}

The behavior of charged polymers has received considerable
attention since the early 70's. However, despite extensive
efforts, much of the phenomena observed in systems containing
polyelectrolytes (PEs) is still not very well understood
\cite{not_understood}. PEs are frequently used in various
industrial applications, such as stabilization of charged
colloidal suspensions and in flocculation processes. They also are
an essential ingredient of many biological systems. These reasons
motivated theoretical
\cite{BarratandJoannyreview,Oosawa,Yamakawa,Dobrynin,deGennes}, experimental
\cite{Baumann,Walczak,Hugel,Hagerman,Spiteri} and computer simulation
\cite{MickaKremer,Winkler,Khan,StevensKremerSimulation} investigations of PEs.
For instance, DNA is known to be a particularly strongly charged
polymer, bearing a charge density of one electron charge per
1.7\AA.

Despite the strong electrostatic repulsion, PEs show a wide range
of complex behavior, depending on the concentration of added salt
and its valency. It is observed that with monovalent counterions,
PEs are usually stretched and assume a rod-like conformation
\cite{Degiorgio,Schmidt,Tricot}. On the other hand, introduction
of a small amount of multivalent counterions significantly reduces
the rigidity of the chain \cite{Baumann,Walczak,Hugel,Hagerman}. Under
certain conditions a PE may completely collapse into a
globule-like conformation \cite{Brilliantov,Solis,Hansen}. For
DNA, this is knows as DNA condensation
\cite{BloomfieldDNAcondensation}.

Even the problem of a single, uniformly charged, polymer in
aqueous solution still poses a theoretical challenge
\cite{Ha1,Ha2,HaPreprint,Ha3,NetzOrland1,BarratJoanny,LiWitten,Odijk,SkolnickFixman,Fixman,LeBret,Shklovskii,Golestanian}.
Single-chain models are much
simpler than real experimental systems, as is the case for
biopolymers in physiological conditions, or with synthetic PEs.
However, much of the interesting phenomena characteristic to
dilute PE solutions is still captured in the models despite the
extended simplification.


The first breakthrough in treating semi-flexible PEs analytically
was made by Odijk \cite{Odijk}, and independently by Skolnick and
Fixman \cite{SkolnickFixman} (OSF) by introducing the concept of
an {\it ``electrostatic persistence length''}.
The notion of persistence length \cite{Yamakawa}, which measures correlations
along the chain,
is very useful in describing elastic properties of
semi-flexible polymers in general, and PEs, in particular.
According to the OSF
theory, the total persistence length of the polymer can be written
as a sum of two contributions: the bare persistence length, $l_0$,
and an electrostatic one, $\ele$. Electrostatic interactions are
treated on a mean-field level and the charges on the polymer are
assumed to be smeared uniformly. Within a linearized version of
the  Poisson-Boltzmann theory, the interaction between any two
charges is screened and given (in units of $\kb T$)
by the Debye-H\"uckel expression,
$\vdh (\br) = \lb \e^{- \kappa r}/r$. The
Bjerrum length $\lb=e^2/ \varepsilon \kb T$
is defined as the distance at which the
electrostatic interaction between two ions of unit charge $e$
equals the thermal energy $\kb T$,
where $\varepsilon$ is the dielectric constant of the medium. For
water at room temperature, $\lb \simeq 7$\AA. The inverse
Debye-H\"uckel screening length is then $\kappa = [4 \pi z (z+1)
\lb c ]^{1/2}$. It depends on the concentration of salt $c$ and
the counterion valency $z$, where throughout this paper we will
explicitly use a $1:z$ salt. Finally, the polymer is considered to
be intrinsically rigid, $l_0 \gg L$. According to the OSF theory,
the electrostatic persistence length then yields
\begin{equation}
    l_{\rm p} = l_0 + l_{\rm OSF} = l_0 + \frac{\lb \lambda^2}{4 z^2 \kappa^2},
\label{OSF}
\end{equation}
where $\lambda=e/a$ denotes the average line charge density along
the chain.

Since the OSF theory is strictly a mean-field theory, the
effective interaction between charges on the polymer is always
repulsive \cite{Fixman,LeBret}. Indeed, Eq.~(\ref{OSF}) indicates
that the polymer becomes more rigid due to electrostatics because
$l_{\rm OSF}>0$. Experiments, however, clearly show that under
some conditions, electrostatics may cause a reversed effect
\cite{Baumann,Walczak,Hugel,BloomfieldDNAcondensation},
where enhanced chain flexibility results from a negative
electrostatic contribution to the persistence length. In order to
account for such behavior, corrections to linearized
Poisson-Boltzmann have been considered
\cite{Fixman,LeBret,Shklovskii,Golestanian}. This was done in two steps. The
first is to take into account the effect of counterion
condensation. With strongly charged polymers, some of the
counterions are loosely bound onto the chain and reduce the
effective charge on the polymer. For a straight infinite cylinder,
this is known as Manning condensation \cite{Oosawa,Manning}. The
second step is to consider correlations between the ions and
thermal fluctuations of the counterion density
\cite{Shklovskii,Golestanian}. Correlations between bound ions
become more significant at lower temperatures. In the $T
\rightarrow 0$ limit, condensed ions are arranged on a periodic
lattice, similar to a Wigner crystal or a strongly correlated
liquid \cite{Shklovskii}. At higher temperatures such correlations
are smeared out due to thermal fluctuations. The latter introduces
another correction to mean-field theory coming from induced
dipoles, similar to van der Waals interactions \cite{Golestanian}.

Both correlations and thermal fluctuations are mechanisms that can
cause the effective interaction between charges on the polymer to
become attractive. Nguyen et al \cite{Shklovskii} considered the
former and calculated the persistence length of a polymer close to
$T=0$. Alternatively, Golestanian et al \cite{Golestanian}
considered the fluctuation mechanism. The two models do not agree
with each other, and there is still no consensus on the question
of which of the mechanisms is more significant at intermediate
temperatures, which are used in experiments
\cite{Diehl,Lau,Kardar,Levin,Nguyen}.

The aim of the present work is to propose a model which takes into
account both correlations and thermal fluctuations
of a single, intrinsically rigid, charged polymer,
immersed in a bulk and continuous dielectric medium. This will
facilitate a closer and consistent examination of the different
mechanisms that cause the fundamentally repulsive electrostatic
repulsion to become attractive. Since the deviation from mean-field
predictions, as seen both in experiments and in previous
theoretical works, is so pronounced, an analytical understanding
of the problem will be of high value.


In the next section, we introduce our model for treating a single,
rod-like PE in presence of added electrolyte (salt) \cite{prl}. In
Section~\ref{meanfield}, a mean-field approximation is used, and
the well known result of OSF is reproduced. Section~\ref{beyondMF}
finds the first-order correction to mean field, taking into
account {\it both} correlations and thermal fluctuations. This is
the main result of the paper where a new expression for the
electrostatic persistence length is obtained. This expression
accounts for the observed attraction between monomers for
strongly charged polymers and multivalent counterions. In the
last sections, our results are compared to experiments and other
theoretical models. Our findings are further discussed in view of these
comparisons.


\section{The Theoretical Model}
\label{model}
\setcounter{equation}{0}

Consider a polymer chain consisting of $N \gg 1$ monomers of
length $a$ each. Taking a worm-like chain approach
\cite{Yamakawa}, the polymer is modeled as a spatial curve $\bR
(s), 0 \le s \le L=N a$, with a total persistence length $\lp$.
Charges on the chain are assumed to be smeared with a positive
constant line charge density $\lambda=e/a$ (one unit
charge $e$ per monomer
size $a$), while mobile ions are taken to be point-like charges.
The system is immersed in a continuous dielectric medium with a
dielectric constant $\varepsilon$. For simplicity, we have assumed
that co-ions are monovalent, while counterions are multivalent,
carrying a charge $-z e$. Namely, the chain is embedded in a $1:z$
electrolyte solution. In order to account for the effect of
counterion (Manning) condensation we follow the two-phase model
introduced by Oosawa \cite{Oosawa}. The first phase is a 1D gas of
counterions that are bound to the polymer and can move only along
the chain. The positions of $I$ bound counterions are denoted as
$\bR (s_1) \dots \bR (s_I)$. The second phase is composed of free
counterions in solution, in equilibrium with the 1D gas. Since the
Manning-Oosawa model regards the PE chains as
an infinite-long cylinder,
we will restrict ourselves hereafter to rod-like polymers which
satisfy, $\lp \gg L$. Finally, we will assume that the effect of
the free ions is to screen out all electrostatic interactions
\cite{StevensKremerSimulation,ZimmLeBret} so that the interaction
between any two charges (smeared charges on the polymer
and bound
$z$-valent counterions) is given by
the screened Debye-H\"uckel  interaction:
$\vdh (\br) = z_i z_j \lb \e^{- \kappa
r}/r$, where $z_i$ and $z_j$ are
the valencies of the two respective ions.
Because we employ a continuum approach, it is necessary to
have $\kappa a \ll 1$ ($a$ is comparable to the size of the
smallest molecule in the system -- free monovalent ions and
monomers). Furthermore, we require that the chain is long enough
so that its contour length, $L$, is much longer than the screening
length, $\kappa L \gg 1$. These limits, $a  \ll \kappa^{-1} \ll L$,
usually hold in experimental and physiological conditions, and are
necessary conditions for our model. In particular, the no-added
salt limit ($\kappa \to 0$) is {\it not} covered by our model.

We can proceed by writing down the grand-canonical partition
function of the system. Up to a normalization constant it is
\begin{equation}
 {\mathcal Z} = \int  {\mathcal D} \bR (s) \left( \sum_{I=0}^{\infty}
    \frac{ \e^{\mu I} }{I !}  \prod_{i=1}^{I}
    \frac{1}{L} \int_0^L \dd s_{i}     \right)
\e^{-H_0 - H_{\rm int}},
\label{partition_function}
\end{equation}
where the path integral is a sum over all possible spatial
conformations of the chain, $\mu$ is the chemical potential of the
1D gas of bound counterions and is related to the counterion
concentration in the bulk reservoir, $H_0$ is the Hamiltonian of a neutral
chain with bare persistence length $l_0$, and $H_{\rm int}$ is the
electrostatic interaction Hamiltonian. It consists of screened
electrostatic interactions between all charged monomers and bound
counterions and is written as a sum of three different
contributions: $H_{\rm int} = H_{\rm mm}+H_{\rm bb}+H_{\rm mb}$,
where
\begin{eqnarray}
\nonumber
   H_{\rm mm} & = & \frac{1}{2} \frac{1}{a^2} \int\limits_0^L \int\limits_0^L \dd s
          \dd s^{\prime}\, \vdh  (\bR (s)-\bR (s^{\prime} ) ),
\\  \nonumber
   H_{\rm mb} &  = & \frac{1}{a} \int\limits_0^L \dd s
             \sum_{i=1}^{I} \vdh ( \bR (s)- \bR ( s_i  ) ),
\\  
   H_{\rm bb} &  = & \frac{1}{2} \sum_{i=1}^{I} \sum_{j=1}^{I}
               \vdh ( \bR ( s_i ) - \bR ( s_j  ) ).
 \label{Hint_Definition}
\end{eqnarray}
All energies are dimensionless and written in terms of the thermal
energy $k_B T$. In this form, the integrations of
Eq.~(\ref{Hint_Definition}) diverge as the terms contain also self
interactions (for instance, when $s \rightarrow s^\prime$ in
$H_{\rm mm}$). All integrations, therefore, should have a lower
cut-off at a distance of order $a$.

In order to treat the interaction term analytically, it is more
convenient to use continuous volume concentrations defined in the
following way \cite{NetzOrland2,NetzReview,Borukhov,Diamant}
\begin{eqnarray}
    \phim ( \br ) &=& \frac{1}{a} \int_0^L \dd s\, \delta (
                   \br - \bR (s) )
\nonumber \\
    \phib ( \br ) &=& \sum_{i=1}^{I}  \delta ( \br - \bR (s_i) ),
\label{phi_def}
\end{eqnarray}
where $\phim$ and $\phib$ are the monomer and bound counterion
concentrations at location $\br$, respectively. These can be
substituted into the partition function,
Eq.~(\ref{partition_function}), by making use of the identity
operator that couples discrete and continuous concentrations. This
is done using the path integral representation of the Dirac delta
function
\begin{eqnarray}
    1 &=& \int {\mathcal D} \phim ( \br ) ~ \delta \left[ \phim ( \br ) -
      \frac{1}{a} \int\limits_0^L \dd s \delta ( \br - \bR (s) ) \right]
\nonumber \\ &=&
        \int {\mathcal D} \phim ( \br ) {\mathcal D}  \psim ( \br )
                        \exp \left\{ - i \int \dd^3 \br
            ~ \psim ( \br ) \right.
\nonumber \\ && \times \left. \left[ \phim ( \br ) -
           \frac{1}{a} \int\limits_0^L \dd s \delta ( \br - \bR (s) ) \right]  \right\}
 \label{psi_Definitions}
\\
  1 &=& \int {\mathcal D} \phib ( \br ) \delta \left[ \phib ( \br ) - \sum_{i=1}^{I}
                     \delta ( \br -  \bR (s_i)  )   \right]
\nonumber \\ &=&
       \int {\mathcal D} \psib  ( \br ) {\mathcal D} \phib ( \br )
      \exp \left\{
       - i \int \dd^3 \br ~ \psib ( \br ) \right.
\nonumber \\
    && \times \left.  \left[     \phib ( \br ) -
       \sum_{i=1}^{I}  \delta ( \br - \bR (s_i) )   \right]  \right\}.
\nonumber
\end{eqnarray}
The extra complexity of this method is the introduction of two new
auxiliary fields, denoted $\psim$ and $\psib$, which couple to
$\phim$ and $\phib$, respectively. Substituting
Eqs.~(\ref{psi_Definitions}) and (\ref{phi_def}) into
Eq.~(\ref{partition_function}), the partition function reads
\begin{eqnarray}
      {\mathcal Z}  &=&  \int {\mathcal D} \bR (s) \left( \prod_{i={\rm m,b}}
      {\mathcal D} \phi^i {\mathcal D} \psi^i \xi_i [\bR] \right)
\nonumber \\
    && \times \exp (-H_{\rm cont}),
\nonumber \\
         \xi_{\rm m} [ \bR ] &=& \exp \left[ -H_{\rm id} + \right.
               \frac{i}{a} \int_0^L \dd s
            \left. \psim ( \bR (s))  + \right.
\nonumber \\
         && +\left.  i \int \dd^3 \br  \phim (\br) \psim (\br) ~ \right]
\nonumber \\
         \xi_{\rm b} [ \bR ] &=& \exp \left\{
           \int \dd^3 \br \left[   a \nb \e^{  i \psib (  \br ) }  \phim ( \br ) \right. \right.
\nonumber \\
      && \left. \left.
       + i \phib (\br) \psib (\br)  \right] \right\}
\nonumber \\
     H_{\rm cont} &=&  \frac{1}{2} \int \int \dd^3 \br \dd^3 \br^{\prime}\,\,
        {\bf \Phi} (\br) \hat{Z}
          {\bf \Phi} (\br^{\prime})  \vdh ( \br- \br^\prime )
\nonumber \\
     {\bf \Phi} &=& { \phim \choose \phib  }  ,~
        \hat{Z}  =
        \left(    \begin{array} {*{3}{c@{\: \:}}c@{\; \;}c}
           1  & -z  \\
           -z  & z^2
                  \end{array}   \right)
\label{cont_z}
\end{eqnarray}
where we have defined
\begin{eqnarray}
    \nb = \e^\mu /L.
\label{nb_def}
\end{eqnarray}
and $\xi_{\rm b}$ was simplified in the following way:
\begin{eqnarray}
    \xi_{\rm b} &=&
    \left( \sum_{I=0}^\infty \frac{1}{I!} \e^{\mu I} \prod_{i=1}^I
             \int\limits_0^L \frac{ \dd s_i}{L} \right)
\nonumber \\
   && \times
    \exp \left[ i \int \dd^3 \br \psib (\br) \left( \sum_i \delta
                   ( \br - \bR(s_i) ) \right) \right]
\nonumber \\
    &=& \left( \sum_{I=0}^\infty \frac{1}{I!} \e^{\mu I} \prod_{i=1}^I
             \int\limits_0^L \frac{ \dd s_i}{L} \right)
    \exp \left[ i \sum_i \psib ( \bR (s_i) ) \right]
\nonumber \\
    &=& \exp \left[ \nb \int\limits_0^L \dd s \e^{ i \psib (\bR(s))} \right]
\nonumber \\
    &=& \exp \left[ a \nb \int \dd^3 \br \e^{ i \psib (\br)} \phim(\br) \right].
\label{explain_xib}
\end{eqnarray}
It is easily seen that carrying out the integrations over the new
fields $\phim,\psim,\phib,\psib$, reproduces
Eq.~(\ref{partition_function}) exactly. However, the form of the
continuous partition function, Eq.~(\ref{cont_z}), is better
organized: single-body interactions of the monomer concentration
$\phim$ and the bound counterion concentration $\phib$ are
collected into the terms $\xi_{\rm m}$ and $\xi_{\rm b}$. The
two-body interaction term $H_{\rm cont}$ has the form of a
quadratic interaction between the concentrations vector field
${\bf \Phi}$, where the interaction between the two vector fields
${\bf \Phi}(\br)$ and ${\bf \Phi}(\br^\prime)$ is given by the
matrix $\hat{Z} \vdh (\br -\br^\prime)$. In the above equation, we
use vector notation to write the interaction between the different
fields as:
\begin{eqnarray}
 &&{\bf \Phi}(\br) \hat{Z} {\bf \Phi}(\br^\prime) \vdh
(\br -\br^\prime)
\nonumber\\
&=&  \left[ \phim (\br) \phim (\br^\prime) -2z
\phim (\br) \phib (\br^\prime)\right.
\nonumber \\
&+& \left. z^2 \phib (\br) \phib (\br^\prime)
\right] \vdh (\br -\br^\prime).
\label{differentPhi}
\end{eqnarray}
This method can be easily generalized in order to account for
any additional species the system may contain, or to different types
of (non-electrostatic) interactions.
For instance, a local interaction can be added to
the matrix elements of $\hat{Z} \vdh (\br -\br^\prime)$ in order
to include  excluded volume interactions.
If counterions are replaced by more
complex charged amphiphiles, a hydrophobic attraction between the
species can be added in a similar manner:
\begin{eqnarray}
    \left(    \begin{array} {*{3}{c@{\: \:}}c@{\; \;}c}
           1  & -z  \\
           -z  & z^2
                  \end{array}   \right)  \vdh (\br -\br^\prime)
   +
    \left(    \begin{array} {*{3}{c@{\: \:}}c@{\; \;}c}
           v_{\rm mm}  & v_{\rm mb}  \\
           v_{\rm mb}  & v_{\rm bb}
                  \end{array}   \right)  \delta(\br-\br^\prime),
\nonumber
\end{eqnarray}
where $v_{ij}$ ($i,j$=m or b) denote second viral
coefficients.

Thus far, the partition function Eq.~(\ref{cont_z}) is exact up to
the general assumptions of the model --- worm-like polymer,
smeared charges on the chain, separation into two phases and
screening by free ions. However, the integrations cannot be
carried out analytically and some approximations have to be made.
The first-order correction just reproduces the well-known
mean-field results, as will be shown below in
Sec.~\ref{meanfield}. Higher-order terms in the expansion
represent corrections to mean field and will be presented in
Sec.~\ref{beyondMF}.


\section{The Mean-Field Approximation}
\label{meanfield}
\setcounter{equation}{0}

\noindent
The approximation method we use is a systematic
expansion in powers of the auxiliary field $\psib$, similar to
loop expansion in field theory \cite{NetzOrland2,NetzReview}.
Expanding to first order in $\psib$ results in
a mean-field approximation. The partition function
takes into account the average interaction between the monomers and the
bound counterions \cite{NetzOrland2}.
Note that this method is somewhat different
than calculating the zeroth-order saddle-point approximation
of the integral over ${\mathcal D} {\bf R}(s)$ in Eq.~(\ref{cont_z}).
For a detailed comparison between the two methods see Ref. \cite{NetzReview}.

Expanding to first order in $\psib$, appearing only in $\xi_{\rm
b}$ we obtain
\begin{eqnarray}
             \xi_{\rm b}  & \simeq &
              \exp \left\{   \int {\rm d}^3 {\bf r}
              \left[ a \nb \left( 1 + i \psib (\br) \right) \phim (\br)
                   \right. \right.
\nonumber \\
  && \left. \left. + i \phib (\br) \psib (\br)  \right] \right\}
\label{gaussian_xib}
\end{eqnarray}
Applying a Fourier transform the partition function reads
\begin{eqnarray}
      {\mathcal Z}_1  &=&  \int {\mathcal D} \bR (s) \left( \prod_{i={\rm m,b}}
      {\mathcal D} \phi^i {\mathcal D} \psi^i \right) \xi_{\rm m} \e^{-H_1}
\nonumber \\
      H_1 &=&  \int \frac{\dd^3 \bk}{(2 \pi)^3} \left[
       \frac{1}{2} {\bf \Phi}^{\dag}_\bk  \hat{Z} {\bf \Phi}_{-\bk} V_\bk^\debh
      + i \phib_\bk \psib_{-\bk} \right.
\nonumber \\ && \left.
      - i a \nb  \phi^{\rm m}_{{\bf k}} \psi^{\rm b}_{ -{\bf k}}
     -  a \nb \phi^{\rm m}_{{\bf k}} (2 \pi)^3 \delta( {\bf k} )
        \right],
\label{z_fourier}
\end{eqnarray}
where ${\mathcal Z}_1$ is the partition function up to first order
in $\psib$. The Fourier transform of  $\phib(\br), \psib(\br)$ and $\phim(\br)$,
is denoted by $\phib_\bk, \psib_\bk$ and $\phim_\bk$, respectively.
The interaction Hamiltonian $H_1$ consists of three
contributions: the first term is the two-body Debye-H\"uckel
interactions. The second term is a bi-linear coupling of the
concentrations fields, $\phim$ and $\phib$, with the auxiliary
field $\psib$, generated by the bound counterions. As for the
third term, we will later show that it is only a constant. The
integrations over the degrees of freedom of the bound ions
$\left\{ \phib,\psib \right\}$ can now be carried out.
\begin{eqnarray}
    {\mathcal Z}_1  &=&  \int {\mathcal D} \bR (s)
      {\mathcal D} \phim {\mathcal D} \psim \xi_{\rm m} \e^{-H_{\rm eff,1}}
\nonumber \\
     H_{\rm eff,1} &=&  \int \frac{\dd^3 \bk}{(2 \pi)^3} \left[  \frac{1}{2} (1-a \nb z )^2
     V_\bk^\debh \phim_\bk \phim_{-\bk} \right.
\nonumber \\ &&
    \left. - a \nb \phim_\bk  (2 \pi)^3 \delta (\bk) \right].
\label{Heff1def}
\end{eqnarray}
The effective interaction Hamiltonian between the monomers,
$H_{\rm eff,1}$, is correct up to first order in $\psib$. The
first term of $H_{\rm eff,1}$ consists of screened electrostatic
interaction ($\vdh$) between Fourier components of the monomer
concentration field $\phim$. The interaction includes a reduced
charge density of $(1-a \nb z )$. Note that $H_{\rm eff,1}$
depends on the conformation of the polymer $\bR (s)$ through the
definition of $\phim$, Eq.~(\ref{phi_def}).

\subsection {Averages and Correlations}

After presenting the partition function, our aim is to
integrate out the degrees of freedom of bound ions and obtain
averaged quantities up to first order in $\psib$. Averages over
configurations of the 1D gas of bound ions are defined as
\begin{eqnarray}
    {\mathcal Z}_1 &=& {\rm Tr}_{ \{ \bR,\phim,\psim \} } \xi_{\rm m} ~
                         {\rm Tr}_{ \{ \phib,\psib \} }   \e^{-H_1}
\nonumber \\
              & = & {\rm Tr}_{ \{ \bR,\phim,\psim \} } \xi_{\rm m} ~  \e^{-H_{\rm eff,1}}
\nonumber \\
    \langle {\cal O} \rangle_1 &=& \frac{ {\rm Tr}_{ \{ \phib,\psib \} } {\cal O} \e^{-H_1} }
                                             {  {\rm Tr}_{ \{ \phib,\psib \} } \e^{-H_1} }
                     =  \frac{ {\rm Tr}_{ \{ \phib,\psib \} } {\cal O} \e^{-H_1} }{ \e^{-H_{\rm eff,1}} } .
\label{z1}
\end{eqnarray}
Taking derivatives of the partition function, different averages
and correlation functions can be calculated. For instance:
\begin{eqnarray}
  \left< \psib_\bk \right>_1 &=& iz(1-a \nb z) \vdh_\bk \phim_\bk
\nonumber \\
  \left< \psib_{\bk_1}  \psib_{\bk_2} \right>_1 &=& \left< \psib_{\bk_1} \right>_1
      \left< \psib_{\bk_2} \right>_1 - (2 \pi)^3 z^2 \vdh_{\bk_1}\delta (\bk_1 + \bk_2)
\nonumber \\
      \left< \phib_{\bk_1}  \phib_{\bk_2} \right>_1 &=& \left< \phib_{\bk_1} \right>_1
      \left< \phib_{\bk_2} \right>_1
\nonumber \\
  \left< \phib_{\bk_1}  \psib_{\bk_2} \right>_1 &=& \left< \phib_{\bk_1} \right>_1
      \left< \psib_{\bk_2} \right>_1 - i (2 \pi)^3 \delta (\bk_1 + \bk_2)
\label{corr1}
\end{eqnarray}

It is interesting to observe that the effective interaction Hamiltonian $H_{\rm eff,1}$
of Eq.~(\ref{Heff1def}) can be rewritten as
\begin{eqnarray}
   H_{\rm eff,1} &=& \int \frac{\dd^3 \bk}{(2 \pi)^3} \left[
        -\frac{i}{2 z} (1-a \nb z) \left< \psib_\bk \right>_1 \phim_{-\bk}
       \right.
\nonumber \\
    && \left. - a \nb \phim_\bk (2 \pi)^2 \delta (\bk) \right].
\end{eqnarray}
Up to a constant prefactor, the first term of $H_{\rm eff,1}$ is
the interaction of the monomer concentration field $\phim$ with
the auxiliary field $\psib$ averaged within mean-field
approximation. Therefore, it is the averaged interaction between
each monomer and the auxiliary field produced by the counterions.

\subsection{Density of bound ions}

In the Manning -- Oosawa's model
\cite{Oosawa,Manning}, the PE is considered as an infinite charged
cylinder, and the average number of bound counterions of valency $z$
per unit length is
\begin{eqnarray}
\nM=\frac{1}{z} \left[ \frac{1}{a} - \frac{1}{z \lb}\right] = \frac{q-1}{z^2 \lb} ,
\label{nM_def}
\end{eqnarray}
where the dimensionless parameter $q=z \lb/a$ is used in Eq. \ref{nM_def}.
Condensation of $z$-valent counterions occurs for $q>1$ and effectively
lowers the value of $q$ to unity, $q_{\rm eff}=1$ \cite{Man_note}.
We are more interested in the case when some of the counterions
are condensed on the polymer. Note that bellow the Manning
condensation threshold ($q<1$) one should simply set $\nM=0$.

The average number of bound counterions per unit length is
\begin{eqnarray}
   n^{\rm tot} [\bR] =\frac{1}{L} \int \dd^3 \br \phib (\br).
\label{ntot_def}
\end{eqnarray}
where the dependence on the polymer conformation, $\bR(s)$, is
through the definition of $\phib$, Eq.~(\ref{phi_def}). Since we
treat here only single polymer chains, the system is assumed to be
infinitely dilute, in the sense that each PE chain occupies only a
small fraction of the overall system volume. Small changes in the
polymer conformation are not expected to change the chemical
potential of the free counterion gas which occupies the entire
volume. As the 1D phase of bound counterions is in equilibrium
with the free counterion phase, their chemical potential is equal.
We conclude that $\mu$, and consequently $\nb = \e^{\mu}/L$,
should not depend on the conformation of the chain in the dilute
polymer limit. Of particular interest is the straight-rod
conformation. In this conformation, the density of bound
counterions should be consistent with the Manning theory, and
$n^{\rm tot} [\bR]$ should, therefore, satisfy
\begin{eqnarray}
   n^{\rm tot} [{\rm rod}] = \nM.
\label{ntot_nm}
\end{eqnarray}
where $\nM$ is the Manning value given in Eq.~(\ref{nM_def}).
Within the mean-field approximation, the average number of bound
counterions per unit length can be obtained by substituting
Eq.~(\ref{phi_def}) and Eq.~(\ref{corr1}) into
Eq.~(\ref{ntot_def})
\begin{eqnarray}
   n_1^{\rm tot} [\bR] &=&
       \frac{1}{L}
   \int \dd^3 \br \left< \phib ({\br}) \right>_1
\nonumber \\
   &=& \frac{1}{L} \int \dd^3 \br \int \frac{ \dd^3 \bk}{(2 \pi)^3} \e^{i \bk \cdot \br}
     a \nb \phim_\bk
\label{nb_1}
\\ & = &
    \frac{a \nb}{L} \int \dd^3 \br \int \frac{ \dd^3 \bk}{(2 \pi)^3} \e^{i \bk \cdot \br}
                  \frac{1}{a} \int\limits_0^L
   \dd s\, \e^{i \bk \cdot \bR(s)}
\nonumber \\
   &=& \frac{\nb}{L}  \int \dd^3 \bk \int\limits_0^L \dd s\, \e^{i \bk \cdot \bR (s)}
        \delta(\bk)
\nonumber \\
  &=&  \frac{1}{L} \nb \int\limits_0^L \dd s = \nb.
\nonumber
\end{eqnarray}
Equation~(\ref{nb_1}) gives us the connection between the average
density of the bound counterions, $n^{\rm tot}[\bR]$, and the
chemical potential, $\mu$ (through $\nb$), up to first order in
$\psib$. According to (\ref{nb_1}), $n_1^{\rm tot} [\bR]$ does not
depend on the conformation of the polymer and is just equal to the
concentration $\nb$. In the following section we will see that
this is strictly a mean-field result. Comparing
Eqs.~(\ref{ntot_nm}) and (\ref{nb_1}) we find that $\nb = \nM$.

\subsection{The Persistence Length}

Substituting $\nb$ into the effective interaction,
Eq.~(\ref{Heff1def}), yields
\begin{eqnarray}
    H_{\rm eff,1} &=& \frac{1}{2} (1-a z \nM)^2 \int \frac{\dd^3 \bk}{(2 \pi)^3}
                 \vdh_\bk \phim_\bk
                 \phim_{-\bk} - L \nM
\nonumber \\
    &=& \frac{L}{2 z^2 \lb} \chi[\bR] - L \nM,
\label{Heff1}
\end{eqnarray}
where
\begin{eqnarray}
   \chi [\bR] &=& \frac{1}{L} \int\limits_0^L \dd s  \int\limits_0^L \dd s^\prime\,
   \frac{ \e^{ - \kappa  | \bR(s) - \bR(s^\prime) | } }{ | \bR(s) - \bR(s^\prime) | }.
\end{eqnarray}
Up to a constant, the integrand is just the screened Coulomb
interaction between any two monomers. The integral is a sum over
all such monomer pairs along the chain with charge density
set at the  Manning value ($q=1$ is equivalent to $\lambda=e/a=e/z
\lb$). This is exactly the interaction Hamiltonian assumed by OSF
for a polymer carrying a uniform line charge density of
$\lambda=e/z \lb$.

Odijk's method for calculating the electrostatic persistence
length \cite{Odijk} assumes small, constant curvature deformations
from the straight rod conformation. The persistence length is then
obtained from the rigidity coefficient of a semi-flexible rod. The
procedure will be described in greater detail in the following
section. Using the effective Hamiltonian $H_{\rm eff,1}$ of
Eq.~(\ref{Heff1}), the OSF result, Eq.~(\ref{OSF}) is reproduced,
with the average density of bound counterions as predicted by
Manning:
\begin{eqnarray}
    l_{\rm p} &=& l_0 + l_{\rm e,1}
\nonumber \\
    l_{\rm e,1} &=&  l_{\rm OSF} = \frac{1}{4 z^2 \kappa^2 \lb}.
\label{le1}
\end{eqnarray}
%


\section{Beyond Mean Field}
\label{beyondMF}
\setcounter{equation}{0}

The results obtained in the previous section are strictly on a
mean-field level \cite{NetzOrland2,NetzReview}. In order to go
beyond this approximation, higher than linear powers of $\psib$
have to be included in the partition function,
Eq.~(\ref{partition_function}). The exact partition function,
Eq.~(\ref{partition_function}), can be rewritten as
\begin{eqnarray}
     {\mathcal Z} &=& {\rm Tr}_{ \{ \bR, \phim, \psim \} } \xi_{\rm m} \e^{-H_{\rm eff,1}}
           \left< \e^{- \Delta H} \right>_1
\nonumber \\
       \Delta H &=& - a \nb \int \dd^3 \br \,\phim (\br) \left[
           \frac{i^2}{2!} \left( \psib (\br) \right)^2 + \dots + \right.
\nonumber \\
   && \left. +
           \frac{i^n}{n!} \left( \psib (\br) \right)^n + \dots \right].
\end{eqnarray}
Performing a cumulant expansion
\begin{eqnarray}
     {\mathcal Z} &=&
       {\rm Tr}_{ \{ \bR, \phim, \psim \} } \xi_{\rm m} \e^{-H_{\rm eff,1}} \exp \left[
            - \left< \Delta H \right>_{\rm c,1} + \right.
\nonumber \\
   && \left. \frac{1}{2!} \left< \Delta H^2 \right>_{\rm c,1}
         - \frac{1}{3!} \left< \Delta H^3 \right>_{\rm c,1}  + \dots
             \right]
\end{eqnarray}
where $\left< {\cal O}^n \right>_{\rm c,1}$ denotes the $n$-th
cumulant. The subscript 1 indicates that the cumulants are
calculated using the first-order expansion of $\Delta H$. For
instance, $\left< {\cal O}^2 \right>_{\rm c,1} = \left( \left<
{\cal O}^2 \right>_1 - \left< {\cal O} \right>_1^2 \right)$, where
the moments $\left< {\cal O}^n \right>_1 $ are defined according
to Eq.~(\ref{z1}).

The effective interaction neglected by $H_{\rm eff,1}$ of
Eq.~(\ref{Heff1}) is therefore
\begin{eqnarray}
         \left< \Delta H \right>_{\rm c,1} - \frac{1}{2!} \left< \Delta H^2 \right>_{\rm c,1}
         + \frac{1}{3!} \left< \Delta H^3 \right>_{\rm c,1}  + \dots
\label{delta_Heff}
\end{eqnarray}
%

\subsection{Second Order Corrections}

\noindent The only term in Eq.~(\ref{delta_Heff}) which is of
second order in $\psib$ is the first term of the first cumulant:
\begin{eqnarray}
        H_2 = \frac{1}{2} a \nb \int \dd^3 {\br} \phim (\br)
              \left< \left[ \psib (\br) \right]^2 \right>_1.
\label{H2}
\end{eqnarray}
Applying a Fourier transform, $H_2$ can be expressed in Fourier
space as well
\begin{eqnarray}
  H_2 &=&
        \frac{1}{2} a \nb \int \frac{\dd^3 \bk_1}{(2 \pi)^3} \frac{\dd^3 \bk_2}{(2 \pi)^3}
         \left< \psib_{\bk_1}
         \psib_{\bk_2} \right>_1 \phim_{-\bk_1-\bk_2}.
\nonumber \\
\label{h2_k_space}
\end{eqnarray}
Substituting $\phim_\bk$ and $\psib_\bk$ into the correlation
expression, Eq.~(\ref{corr1}), yields a correction to the
effective interaction Hamiltonian, $H_{\rm eff,1}$, obtained in
Eq.~(\ref{Heff1})
\begin{eqnarray}
        H_{\rm eff,2} &=& -\frac{1}{2} z^2 \frac{\nb \lb^2}{a^2} (1-a z \nb)^2
            \left[\int\limits_0^L \dd s_0 \int\limits_0^L \dd s_1  \int\limits_0^L \dd s_2\right.
\nonumber \\
   && \times\left.
      \frac{ \e^{ - \kappa  | \bR(s_1) - \bR(s_0) | } }{ | \bR(s_1) - \bR(s_0) | }
     \frac{ \e^{ - \kappa  | \bR(s_2) - \bR(s_0) | } }{ | \bR(s_2) - \bR(s_0) | }\right]
\nonumber \\
      && + ~~z^2 \lb \nb N.
\label{Heff2}
\end{eqnarray}
The above result is obtained by using a lower cut-off at distance
$a$ on one of the three integrations, and expanding the integral
in powers of $\kappa a$, neglecting all but the leading term. In
the expansion method, $\psib$ represents an auxiliary field that
is generated by the bound counterions
\cite{NetzOrland1,NetzOrland2}. Examining the mean-field
interaction, Eq.~(\ref{h2_k_space}), each mode of the monomer
concentration field $\phim$ interacts with the average of two
$\psib$ auxiliary fields. The result is a
counterion-monomer-counterion interaction, and appropriately, the
first term of $H_{\rm eff,2}$ in Eq.~(\ref{Heff2}) has the form of
a three-body interaction. As explained in the previous section,
the chemical potential $\mu$ does not depend on the polymer
conformation $\bR(s)$. As a consequence, the second term of
$H_{\rm eff,2}$ in Eq.~(\ref{Heff2}) does not depend on the
conformation $\bR(s)$ and does not contribute to the persistence
length.

\subsection{Density of Bound Ions}

Taking into account second-order corrections, the
average number of bound counterions changes as well
\begin{eqnarray}
  n_2^{\rm tot} [\bR] &=& \frac{1}{L} \int \dd^3 \left< \phib (\br) \right>_2
\nonumber \\
   &=& \nb
    + \nb z (1-a z \nb) \frac{\lb}{a} \chi [\bR],
\label{average_density2}
\end{eqnarray}
where $\left< {\cal O} \right>_2$ denoted the average of ${\cal
O}$ calculated with the second-order Hamiltonian, $H_2$ of
Eq.~(\ref{h2_k_space}).

According to the theory of Manning condensation, the density of
bound counterions for a straight rod is $n_2^{\rm tot} [{\rm
rod}]= \nM$ \cite{Man_note,HaLiu}. The chemical potential $\mu$,
defined through $\nb$, should therefore satisfy
\begin{eqnarray}
     a \nb + \nb z (1-a z \nb) \lb \chi[{\rm rod}] = a \nM.
\label{eq_for_nb}
\end{eqnarray}
where $\chi[{\rm rod}]$ is the value of $\chi [\bR]$ for the
straight rod conformation, in which $| \bR(s) - \bR(s^\prime) | =
|s-s^\prime|$. Substituting in Eq.~(\ref{eq_for_nb}) the
expression for $\nM$, Eq.~(\ref{nM_def}), yields
\begin{eqnarray}
     \nb = \frac{ q \chi[{\rm rod}] + 1 \pm \sqrt{ (q \chi[{\rm rod}] - 1)^2 + 4 \chi[{\rm rod}] } }
                     { 2 z^2 \lb \chi[{\rm rod}] }.
\label{nb2}
\end{eqnarray}
In the limit of extremely high salt concentrations, $\kappa a \gg
1$, all correlation and fluctuation effects are screened out
completely and decay exponentially with $\kappa a$. The smaller of
the two solutions of Eq.~(\ref{eq_for_nb}) is therefore chosen in
order that the Manning value $n_2^{\rm tot} [\bR] =\nM$ is
reproduced: $\nb[ \kappa a \rightarrow \infty ] = \nM$. As
explained in the previous section, the chemical potential $\mu$
(and consequently $\nb= \e^{\mu}/L$ is not expected to depend
on the polymer conformation. The density of the bound counterions
$n_2^{\rm tot} [\bR]$ does, however, depend on the polymer
conformation through Eq.~(\ref{average_density2}).

\subsection{The Persistence Length}

Using the effective Hamiltonian, Eq.~(\ref{Heff2}), and
the chemical potential, Eq.~(\ref{nb2}), it is now possible to
repeat Odijk's method and calculate the electrostatic persistence
length. The difference between the effective interaction energy at
a general conformation $\bR(s)$ as compared to the straight rod
one is
\begin{eqnarray}
      \Delta H_{\rm eff} &=& H_{\rm eff,1} [\bR] + H_{\rm eff,2} [\bR]
         -
\nonumber \\
&& H_{\rm eff,1} [{\rm rod}] - H_{\rm eff,2} [{\rm rod}].
\label{deltaHeff}
\end{eqnarray}
Odijk's method for calculating the persistence length requires
expanding $\Delta H_{\rm eff}$ in small deformations of the chain
around the straight rod conformation \cite{Odijk}. In this limit,
the distance between points $s$ and $s^\prime$ can be approximated
as
\begin{eqnarray}
      | \bR(s) - \bR(s^\prime)| &\simeq&  | s-s^\prime | \left[ 1-\alpha(s,s^\prime) \right] + O(\alpha^2)
\nonumber \\
      \alpha (s,s^\prime) &=& \frac{1}{24} \left( \frac{s-s^\prime}{\rho} \right)^2,
\label{odijk_approximations2}
\end{eqnarray}
where $\rho \gg L$ is the small overall radius of curvature of the
fluctuating chain (not to be confused with a spontaneous radius of curvature).
The persistence length is then given by
\begin{equation}
    l_{\rm e,2}=  2 \frac{\rho^2}{L} \Delta H_{\rm eff}.
\label{persistence_length_general}
\end{equation}
It is, therefore, additive in terms of $\Delta H_{\rm eff}$:
\begin{eqnarray}
    l_{\rm e,2} &=&  2 \frac{\rho^2}{L} \left( H_{\rm eff,1} [\bR] -
    H_{\rm eff,1} [{\rm rod}] \right) +
\nonumber \\
  && + 2 \frac{\rho^2}{L}
    \left( H_{\rm eff,2} [\bR] - H_{\rm eff,2} [{\rm rod}]
    \right).
\label{persistence_length_e2}
\end{eqnarray}
On the other hand, we note that the electrostatic persistence
length is {\it not} additive in orders of $\psib$ ($ l_{\rm e,1}
\neq 2 \frac{\rho^2}{L} ( H_{\rm eff,1} [\bR] - H_{\rm eff,1}
[{\rm rod}])$ ) as the expression for the chemical potential also
changes, as compared to the mean-field approximation. Inserting in
Eq.~(\ref{persistence_length_e2}) the value for $\nb$,
Eq.~(\ref{nb2}), and expanding the result in powers of $1/\rho$
the different integrations can be evaluated. This requires cutting
of all ultra-violet divergences at a distance $a$. In the limit of
$a \ll \kappa^{-1} \ll L$ we find
\begin{eqnarray}
      \lp &=& l_0 + l_{\rm e,2}
\nonumber \\
      l_{\rm e,2} &=& l_{\rm OSF}
          \left[ q (2-q) - \frac{ (q-1)^2 }{q \ln \kappa a}
          \right],
\label{le2}
\end{eqnarray}
where we have expanded $l_{\rm e,2}$ in $(\ln \kappa a)^{-1}$ and
kept the two leading terms. Note that we have already taken the
first order (mean field) interaction into account in
Eq.~(\ref{deltaHeff}). Equation~(\ref{le2}) is our main prediction
and is depicted in Fig.~1 for different counterion valencies
$z=1,2,3$ as a function of $\kappa a$. At low salt concentrations
($\kappa a\ll 1$) or high $q$,  the persistence length maintains
the OSF $\kappa^{-2}$ dependence, $l_\e\sim l_{\rm OSF}\sim
\kappa^{-2}$. We find that the electrostatic persistence length
depends strongly on the valency of the counterions. For monovalent
counterions, $l_\e$ is usually positive, indicating an effective
repulsion between the monomers. However, its value is smaller than
the one predicted by OSF. Introduction of multivalent counterions
reduces significantly the rigidity of the PE  and usually $l_\e<0$,
indicating an effective attraction between monomers.

The vanishing of the total persistence length $\lp$ under certain
conditions is alluding to the phenomena of PE collapse. A full
consideration of the rod-globule transition requires a more
consistent elastic model for the polymer chain than the
persistence length prescription used here. Furthermore, Odijk's
method for calculating $l_\e$ does not hold for flexible polymers
\cite{BarratJoanny}. However, the condition that total persistence
length vanishes, $\lp=0$, is the validity limit of the rod-like
regime and is indicative of some mechanical instability. For
instance, using parameters applicable to DNA chains: $l_0=500$\AA,
3:1 salt, $a=1.7$\AA, and $l_B=7$\AA\ we get a DNA collapse at
$\kappa^{-1}\simeq 30$\AA, corresponding to a 3:1 salt
concentration of about 17m\,M.

Expanding the density of the bound ions,
Eq.~(\ref{average_density2}), for small deformations of the chain
yields
\begin{eqnarray}
   n^{\rm tot}_2 [\rho]= \nM \left( 1 - \frac{1}{ 8 \kappa^2 \ln \kappa a }
       \frac{1}{\rho^2} \right),
\label{n_tot}
\end{eqnarray}
For a straight chain, $\rho \rightarrow \infty$ and the Manning
value $\nM$ is obtained. However, more counterions are condensed
on a bent polymer with a finite
radius of curvature. This enhanced  condensation drives
a further reduction of the persistence length. For a rod-like
polymer, $\rho^2 = L \lp/3$ and the correction to $\nM$ is of
order $1/N$. This difference does, however, have a significant
effect on the electrostatic persistence length, because $H_{\rm
eff,2}$ is a triple integral over the monomers. In order to
examine the effect of increased condensation, we look at the
asymptotic form of Eq.~(\ref{le2}) for $q = 1+ \Delta q, \Delta q
\ll 1$, in two cases. In the first we allow the density of the
bound counterions to be adjusted according to the equilibrium
condition with the bulk (this is an expansion of Eq.~(\ref{le2})
in power of $\Delta q$). This is consistent with the general
considerations of our paper. In the second case we add a
constraint that fixes the density to be according to the Manning
theory for all conformations of the polymer. This more restrictive
constraint is added in order to make comparisons with other
models. Expanding in $\Delta q$ we recalculate $l_\e$ for both
cases
\begin{eqnarray}
    l_\e &=& l_{\rm OSF} \left[
       1 + {\rm O} ( \Delta q^2 )
        \right] \\
   l_\e^{\rm fixed} &=& l_{\rm OSF} \left[
       1 - [1/ \ln (\kappa a)] \Delta q + {\rm O} ( \Delta q^2 )
        \right]. \nonumber
\label{le2_restricted}
\end{eqnarray}
The lack of a linear term in $\Delta q$ in the first expression of
Eq.~(\ref{le2_restricted}) indicates that corrections to Manning
condensation for bent polymer chains have a substantial influence
on the persistence length. Fig.~2 depicts the excess number of
bound counterions as a function of $q$ for three salt
concentrations (corresponding to $\kappa a = $0.02, 0.04 and 0.08).

An interesting effect is charge inversion, where the total charge
of the polymer with the bound counter-ions changes sign ($z n^{\rm
tot} < 1/a$). The persistence length at which charge inversion
occurs is given by
\begin{eqnarray}
    l_{\rm p}^{\rm inv} = \frac{ 3 z \nM }{ 8 \kappa^2 (a z \nM-1) \ln \kappa a}
    \frac{1}{N}.
\end{eqnarray}
According to our model, charge inversion will not occur on a long
($N \gg 1$), rod-like polymer, because $l_{\rm p}^{\rm inv}
\propto 1/N$.

\subsection{Higher Order Corrections}

 In order to examine the convergence of the loop expansion
used above, we have calculated the next two orders of
approximation: $(\psib)^3$ and $(\psib)^4$. The effective
interactions are obtained by taking into account terms of higher
orders of $\psib$ in Eq.~(\ref{delta_Heff}). The 3rd and 4th order terms
are given as:
\begin{eqnarray}
   H_3 &=& - a \nb \int \dd^3 \br \phim (\br) \frac{i^3}{6}
         \left<  \left[ \psib (\br) \right]^3 \right>_1
\nonumber \\
       H_4 &=& - a \nb \int \dd^3 \br \phim (\br) \frac{i^4}{4!}
    \left< \left( \psib (\br) \right)^4 \right>_1
\nonumber \\
    && - \frac{1}{2} a^2 (\nb)^2 \left\{ \left< \left[ \int \dd^3 \br \phim
    (\br) \frac{i^2}{2} \left( \psib (\br) \right)^2
    \right]^2 \right>_1  \right.
\nonumber \\
   && \left. - \left< \int \dd^3 \br \phim
    (\br) \frac{i^2}{2} \left( \psib (\br) \right)^2  \right>_1^2  \right\}.
\nonumber \\
\label{h34_def}
\end{eqnarray}
Higher orders of $\psib$ can be taken into account following the
same prescription used above for calculating the chemical
potential and the persistence length for the second-order
correction. The expression for $l_{\rm e,3}$ and $l_{\rm e,4}$ are
not detailed here because they are quite complex. However, they do not
change the polymer behavior in any qualitative fashion. We find
that Eq.~(\ref{le2}), valid to second order, accounts for most of
the deviation from the OSF result. Third and fourth order terms
represent only a relatively small correction to the second-order
one. For instance, for $q>10$ and $\kappa a < 0.01$, the
third-order correction is less that 4\% of the second-order one.
The fourth-order correction is again less than 4\% than the
third-order one.

Furthermore, we find that the convergence is better for large $q$
and small $\kappa a$ \cite{NetzReview}.
This is the more interesting $q$ regime since, for instance,
in DNA solutions with trivalent
counter-ions we get roughly $q \simeq 12$.


\section{Comparison with Other Models}
\label{Models}
\setcounter{equation}{0}

\noindent
Our model is closely related to several previous
ones. Ha and Thirumalai have used a similar loop expansion method for
calculating the persistence length of polyampholytes \cite{Ha3}
and of bundles of polymers \cite{HaPreprint}. For the case of a
single PE, Golestanian et al \cite{Golestanian} took into
account thermal fluctuations of the bound counterions density.
This is strictly an all-fluctuations model, which is expected to
become accurate at high temperatures ($q = 1+\Delta q, \Delta q
\ll 1$). Their expression for the electrostatic persistence
length is \cite{Golestanian}
\begin{eqnarray}
    l_{\rm e}^{\rm fluct} &=& \frac{\lb}
        {4 q^2 \kappa^2 a^2 \left[ 1 -(q-1) \ln (\kappa a)
        \right]^2},
\label{le_fluct}
\end{eqnarray}
where we have explicitly omitted terms with a higher order dependence on
$\kappa a$.

A second model, suggested by Nguyen et al \cite{Shklovskii}
assumes that condensed ions are arranged in a Wigner crystal, or a
strongly correlated liquid on an infinite cylinder with diameter
$d$.
This picture becomes accurate at low temperatures ($q
\gg 1$).
For the case of no-added salt, their expression for the electrostatic
persistence length is:
\begin{eqnarray}
    l_{\rm e}^{\rm corr} = \lb \sqrt{z} ( d/a )^{3/2}.
\end{eqnarray}
As Nguyen et al did not take into account the effect of
salt  (they considered only counterions), we
have slightly modified their derivation for the purpose of
comparison with our model.
Equation~(13) of Ref. \cite{Shklovskii}
estimates the interaction energy of an ion with its Wigner-Seitz
cell of background charges. Up to a constant of order unity,
it is found to be
\begin{eqnarray}
    \epsilon (n) \simeq - \frac{n^{1/2} z^2 e^2 }{\varepsilon}
\label{epsilon_shklovskii}
\end{eqnarray}
where $n$ is the average surface charge density on the cylinder.
The only
change we include in the above  equation is to assume that in the
presence of salt, the interaction energy should be proportional
to the number of Wigner-Seitz cells that reside within a circle
with a radius equal, roughly, to the screening length $\kappa^{-1}$.
Up to a constant prefactor, this modification gives
\begin{eqnarray}
    \tilde{\epsilon} (n) \simeq -\frac{ n^{1/2} z^2 e^2 }{\varepsilon}
    \frac{n}{\kappa^2},
\label{epsilon_ours}
\end{eqnarray}
Repeating  the
derivations of Ref. \cite{Shklovskii} with this modification of their Eq.~(13),
and assuming that $d \simeq a$ yields
\begin{eqnarray}
    l_{\rm e}^{\rm corr} &\simeq& - l_{\rm OSF} \frac{q^2}{\sqrt{z}}.
\label{le_corr}
\end{eqnarray}

Our loop expansion method can be shown to account qualitatively for both
limits of the parameter $q$.
{As discussed in the previous section, we find $q$ to be
the relevant, temperature dependent parameter that determines the
system behavior.
This is the reason we expand the results
in the two limits $q \gg 1$ and $q \gtrsim 1$ rather than
low or high temperatures ($\kappa \rightarrow \infty$ or $0$).
The limits of very low or high temperatures
are beyond the validity range of our model which explicitly assumes that
$L^{-1} \ll \kappa \ll a^{-1}$.
However, experimental systems usually have
a large value of $q$ with a finite screening length.
For instance, typical parameter
values in experiments with DNA segments at room temperature
are $l_{\rm B} =7$\AA, $a=1.7$\AA, $\kappa^{-1}=10-100$\AA~and
$N \geq 150$. With tri-valent counterions ($z=3$) we get
$q \simeq 12 \gg 1$ and $\kappa L>1$.
This is an example where we can consider the relatively high $q$
limit at room temperature.

In order to compare the three models, we expand the
expressions for the electrostatic persistence length:
ours $l_{\rm e}^{\rm loop}$; $l_{\rm e}^{\rm
fluct}$ of Ref. \cite{Golestanian}; and $l_{\rm e}^{\rm corr}$ of
Ref. \cite{Shklovskii}, in these two limits.

Including terms up to linear order
in $\Delta q$ close to $q=1$ we get
\begin{eqnarray}
    l_{\rm e}^{\rm fluct} &=& l_{\rm OSF} \left[
        1 - 2 \ln (\kappa a) \Delta q + {\rm O} ( \Delta q^2 )
        \right]
\nonumber \\
    l_{\rm e}^{\rm corr} &\simeq& l_{\rm OSF} \left[
       -1 / \sqrt{z} - 2 / \sqrt{z} \Delta q + {\rm O} ( \Delta q^2 )
        \right]
\nonumber \\
    l_{\rm e}^{\rm loop} &=& l_{\rm OSF} \left[
       1 + {\rm O} ( \Delta q^2 )
        \right].
\label{compare_small_q}
\end{eqnarray}
However, the calculations of Golestanian et al assumes that the
amount of condensed counterions is according to Manning for all
polymer conformations. In the loop calculation, we have relaxed
this assumption and took the Manning counterion value only for
the completely straight rod case. For the sake of comparison, we
impose now this restriction. This has been done already in
Eq.~(\ref{le2_restricted})
\begin{eqnarray}
   l_{\rm e}^{\rm loop,fixed} & = & l_{\rm OSF} \left[
       1 - \Delta q / \ln (\kappa a)  + {\rm O} ( \Delta q^2 )
        \right] .
\end{eqnarray}
With the new restriction, the linear term in $\Delta q$ reappears
but with a different coefficient than in $l_{\rm e}^{\rm corr}$ of
Eq.~(\ref{compare_small_q}). The differences in the coefficients,
as well as the different dependence on the cut-off distance $a$ is
due to the different methods and approximations used
in calculating the persistence length.

The second case is that of large $q$, for which the models give:
\begin{eqnarray}
    l_{\rm e}^{\rm fluct} &= & l_{\rm OSF} \left[
         1/[q \ln (\kappa a)]^2 + {\rm O} ( 1/q^3 )
        \right]
\nonumber \\
    l_{\rm e}^{\rm corr} & = & l_{\rm OSF} \left[
       - q^2 / \sqrt{z}  + {\rm O} ( q ) \right]
\nonumber \\
    l_{\rm e}^{\rm loop} &=& l_{\rm OSF} \left[
        - q^2 + {\rm O} ( q ) \right].
\end{eqnarray}
We note that the electrostatic persistence length
of the two previous
models \cite{Golestanian,Shklovskii} depends very differently
on each of the fundamental
parameters of the system: the charge density of the polymer
$\lb/a$, and the valency of the counterions $z$.
Furthermore, their expressions do not have similar limits
in the two $q$ limits discussed above.
However, in the limit $q \gtrsim 1$, our result is similar
to the one obtained by the fluctuation model \cite{Golestanian}.
Conversely, in the limit $q \gg 1$, our result resembles the one
obtained by the correlation model \cite{Shklovskii}.
Some discrepancies are apparent.
As explained above,
the difference between our model and the fluctuations one is
mainly due to the different method used for obtaining the
persistence length. The difference with the correlations
governed model is mostly due to the discreteness of the charges
assumed in Ref. \cite{Shklovskii} and their specific arrangement
in a 2D Wigner lattice.

Our expression for the electrostatic persistence, Eq.~(\ref{le2}),
neither vanishes no diverges in the limits or low or high temperatures.
Fluctuation contributions to the electrostatic persistence length
vanish in the limit $T \rightarrow 0$. Conversely,
correlation contributions vanish in the limit
$T \rightarrow \infty$.
This illustrates that Eq.~(\ref{le2}) contains contributions from both
fluctuations and correlations.


\section{Comparison with Experiments}
\label{Experiments}
\setcounter{equation}{0}

\noindent Comparison between our expression for the persistence
length, Eq.~(\ref{le2}), with that obtained in experiments
\cite{Baumann,Walczak,Hugel,Hagerman,Spiteri} is difficult. Although $l_{\rm e,2}$
correctly predicts that the persistence length should be smaller
than OSF, it seems that the reduction we obtain is too large as
compared with experiments. Actually, at present, we are not aware
of any other theoretical modeling which explains quantitatively
the experimental data.

Measurements of rigid PEs usually involve short DNA segments.
Experiments show that adding of very small amounts of
multivalent counterions greatly reduce the persistence length of
DNA, $l_{\rm p}=l_0+l_{\rm e}$ even bellow its bare value $l_0$
\cite{Baumann,Hagerman,Raspaud}, indicating a negative
electrostatic persistence length. However, substituting DNA
parameters ($a \simeq 1.7$\AA, $l_0 \simeq 500$\AA) and salt
concentrations common to experiments ($10$\AA\
$<\kappa^{-1}<100$\AA\ ) into the total persistence length, $l_{\rm
p}$, of Eq.~(\ref{le2}),  we find  that DNA should  collapse ($l_{\rm
p}<0$), for $z=3$ or 4. This does not agree with experiments
where the DNA is still in the rigid-rod limit for the same system
parameters.

This discrepancy
may be caused by several important experimental features which are
neglected in our model as well as in Refs. \cite{Golestanian,Shklovskii}.
DNA segments are prepared in a buffer which stabilizes the solution pH
and removes free divalent calcium ions  \cite{Raspaud}. The buffer itself contributes a
finite, non-negligible concentration of monovalent ions.
In a second stage,  multivalent ions such as
spermidine ($z$=3) or spermine ($z$=4) are added.
In experiments, the monovalent salt concentration may be
much higher than the multivalent one,  making
the contribution of the multivalent salt to the screening length
$\kappa^{-1}$ quite negligible. In the model we did not take into account
mixtures of mono- and multi-valent ions.

According to the Manning-Oosawa model which was employed by us,
entropy considerations
dictate that condensation of multivalent counterions is much
favorable than monovalent ones. At low concentrations of
multivalent salts, experiments clearly show that this is not
always the case \cite{Raspaud,Pelta,Saminathan}. At low but finite
polymer and multivalent salt concentrations (as is usually the
case in experiments), entropy and the finite size of the
counterions (spermidine and spermine are relatively large
molecules) prohibit multivalent counterions from condensing on the
chain. Generalization of the
Manning-Oosawa model to account for this effect is not
straightforward.

In experiments the change from a rigid rod-like behavior to a flexible
one is highly sensitive to the multivalent concentration
 \cite{Baumann,Hagerman}.  The above discussion emphasizing
the deviation from the Manning-Oosawa model may also explain
this  change-over.
At low concentrations of multivalent
ions, less counterions are condensed than according to the Manning value.
Hence, the chain is still rigid in disagreement with our prediction.
At higher multivalent
salt and polymer concentrations, where the Manning value for
condensation on a cylinder $\nM$ holds, the DNA is completely
collapsed, making both theory and experimental measurements of
persistence length useless.  Therefore, the main difficulty
in  our (and similar) models is the
small window of parameter values where the
model can be applied. Experiments with DNA do not, in general,
fall in this window due to the strong charging of the chains.

We briefly  mention other features not considered in our model,  and which may
influence the persistence length. They include the finite size of the
counterions \cite{ci_size}, the ordering of the charges along the
polymer chain \cite{charge_order,GC}, the concentration profile of
the condensed ions around the polymer \cite{ZimmLeBret} and other,
more specific, details of the polymer type and ions used in
experiments \cite{Bloomfield}.


\section{Summary}
\label{Summary}
\setcounter{equation}{0}

\noindent We have found significant corrections to the persistence
length of a single, stiff, strongly charged and long PE, as
compared to the standard mean-field result of Odijk-Skolnick-Fixman.
Our method takes into account
both correlations between condensed ions and thermal fluctuations.
At low salt concentrations, the calculated electrostatic
persistence length $l_{\e,2}$ is proportional to $l_{\rm
OSF}$. However, the prefactor, which depends on $q=z\lb/a$,
drastically changes the system behavior. For $q \le
1$, $l_{\rm e,2} = l_{\rm OSF}$ is obtained exactly. For $1<q<2$, the
electrostatic persistence length $l_{\e,2}$ is positive,
indicating an effective repulsion between the monomers. For $q>2$,
the interaction becomes attractive causing a reduction in the
chain stiffness, $l_{\rm e,2}<0$. This observation is in agreement
with the reduction in persistent length observed in experiments
with multivalent counterions and strongly charged polymers.

We compared our result for the electrostatic persistence length,
Eq.~(\ref{le2}), with two previous models and found that our model
takes into account both thermal fluctuations and correlations between
bound counterions. Our model qualitatively agrees with both
previous ones at different limits of the parameter $q$.

It is interesting to note that $q=2$ corresponds to the case where the
average electrostatic interaction between bound ions equals $\kb T$.
This means that for $1<q<2$, thermal fluctuations are expected
to dominate over correlations. On the other hand, for $q>2$,
correlations become more significant. Our conclusion is  that although
thermal fluctuations reduce the (mean field) electrostatic
repulsion between monomers, they are not sufficient
to induce effective attraction. In order to correctly describe the
attractive, collapsed case, correlations between counterions have
to be included.

We have also obtained the average density of bound ions and found
that more counterions condense on the chain than is predicted by
the Manning-Oosawa model. The increased condensation has a
significant effect on the persistence length and cannot be
neglected. Furthermore, we have estimated the conditions under which
collapse of a rigid PE may occur. The results are reasonable and
relate, at least qualitatively, to the phenomena of DNA
condensation. As explained in previous sections, our theory cannot yet be
directly compared with experimental results (in particular with DNA).

We believe that additional work is
needed to shed more light on the mechanical instability of the
chain, indicating a rod-globule transition. Different, more complex, methods
are required in order
to calculate the persistence length of flexible chains. For
instance, the validity range of some variational methods are known
to be wider than that of the OSF theory
\cite{NetzOrland1,BarratJoanny}, and may better apply to flexible
chains. Moreover, the entire
notion of persistence length for describing the chain elastic
properties breaks down close to the instability. The vanishing of
the persistence length indicates that the method is no longer
consistent, and higher powers of the radius of local curvature
$\partial^2 \bR/\partial s^2$ have to be taken into account.

DNA experiments are usually performed for polymer
and salt concentrations requiring a more detailed examination of
the counterion condensation phenomena than the simplified
Manning-Oosawa model. Some effects that are unique to mixtures of
different types of ions need to be taken into account. A
quantitative analysis of this phenomenon requires further and
rather detailed considerations and will be presented elsewhere.


\vspace{1cm}
\noindent
{\em Acknowledgments}

We would like to thank I. Borukhov,  H. Diamant, A.
Grosberg, B.-Y. Ha, M. Kozlov, R. Mints, R. Netz, T. Odijk, H. Orland,
Y. Rabin, B. Shklovskii, D. Thirumalai, T. Witten
and, in particular, to Y. Burak for
useful discussions and correspondence. Partial support from the
U.S.-Israel Binational Science Foundation (B.S.F.) under grant No.
98-00429 and from the Israel Science Foundation founded by the Israel
Academy of Sciences and Humanities
--- centers of Excellence Program, and under grant No. {210/02} is gratefully
acknowledged.

\pagebreak


\section*{Figure Captions}

\noindent {\bf Fig.~1}:
  The electrostatic persistence length $l_{\rm e}$ as function of $\kappa a$ according
to OSF (dashed line) and our $l_{\rm e,2}$ of Eq.~(\ref{le2})
(solid line). Valencies are specified next to each curve. The
parameters chosen are: $a=4$\AA, $\lb=7$\AA , so that $q=1.75 z$.
The negative $l_\e$ values for $z=2, 3$ indicate a possible
collapse transition of the PE chain.

\vspace{1cm}

\noindent {\bf Fig.~2}: The excess number of bound counterions
$L(n^{\rm tot}_2 [\rho]-\nM)$ according to Eq.~(\ref{n_tot}), as a
function of $q$. The number of excess bound ions is plotted for
three salt concentrations corresponding to $\kappa a=$0.02, 0.04
and 0.08. The radius of curvature was calculated according to
$l_{\rm e,2}$, Eq.~(\ref{le2}), with $z=3, a=4$\AA\ and
$l_0=500$\AA.

\newpage

\begin{figure}[tbh]
\epsfxsize=1.0\linewidth
\centerline{\hbox{ \epsffile{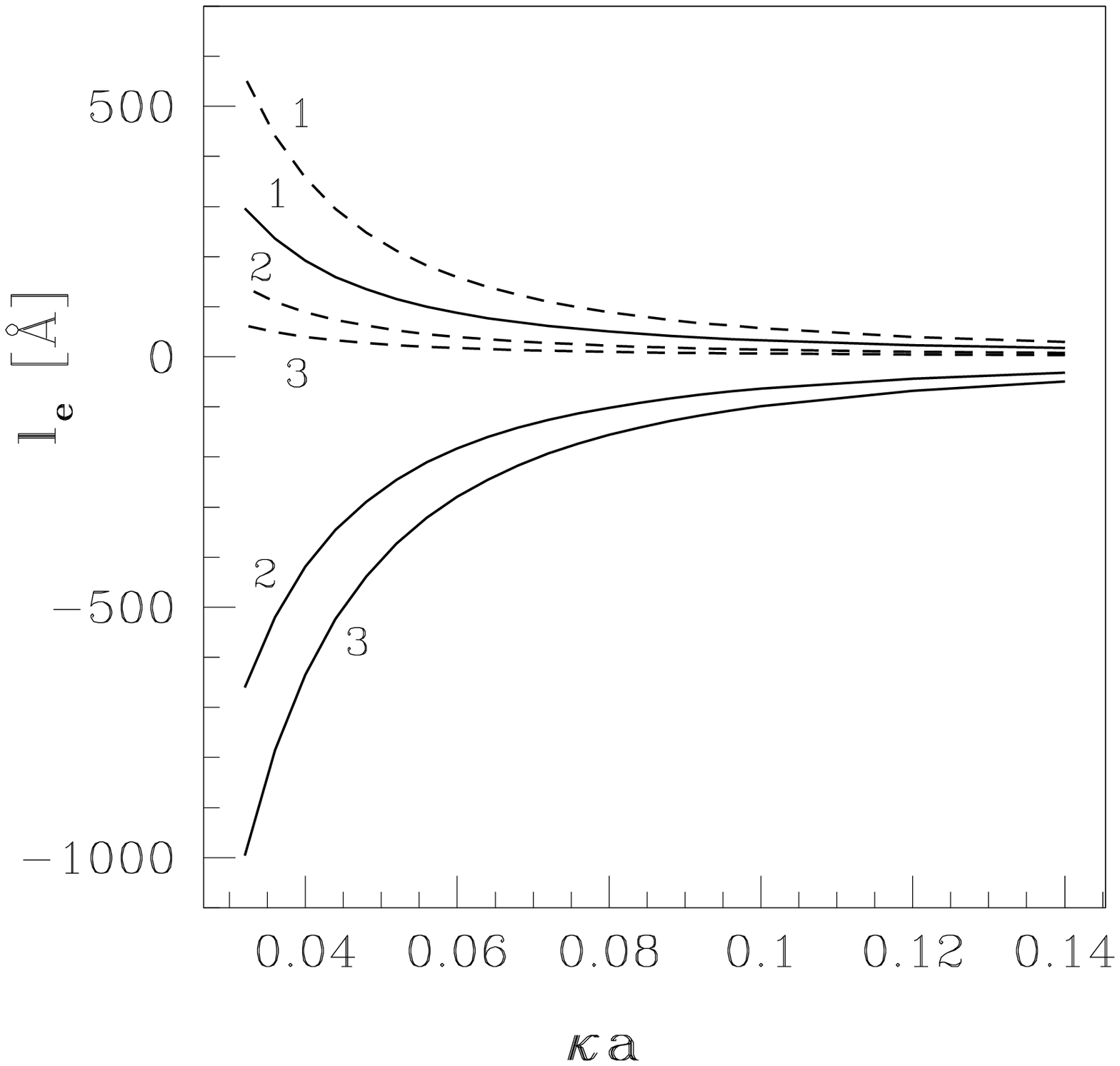} } }
\end{figure}
\centerline{\large Fig.~1}

\newpage

\begin{figure}[tbh]
\epsfxsize=1.0\linewidth
\centerline{\hbox{ \epsffile{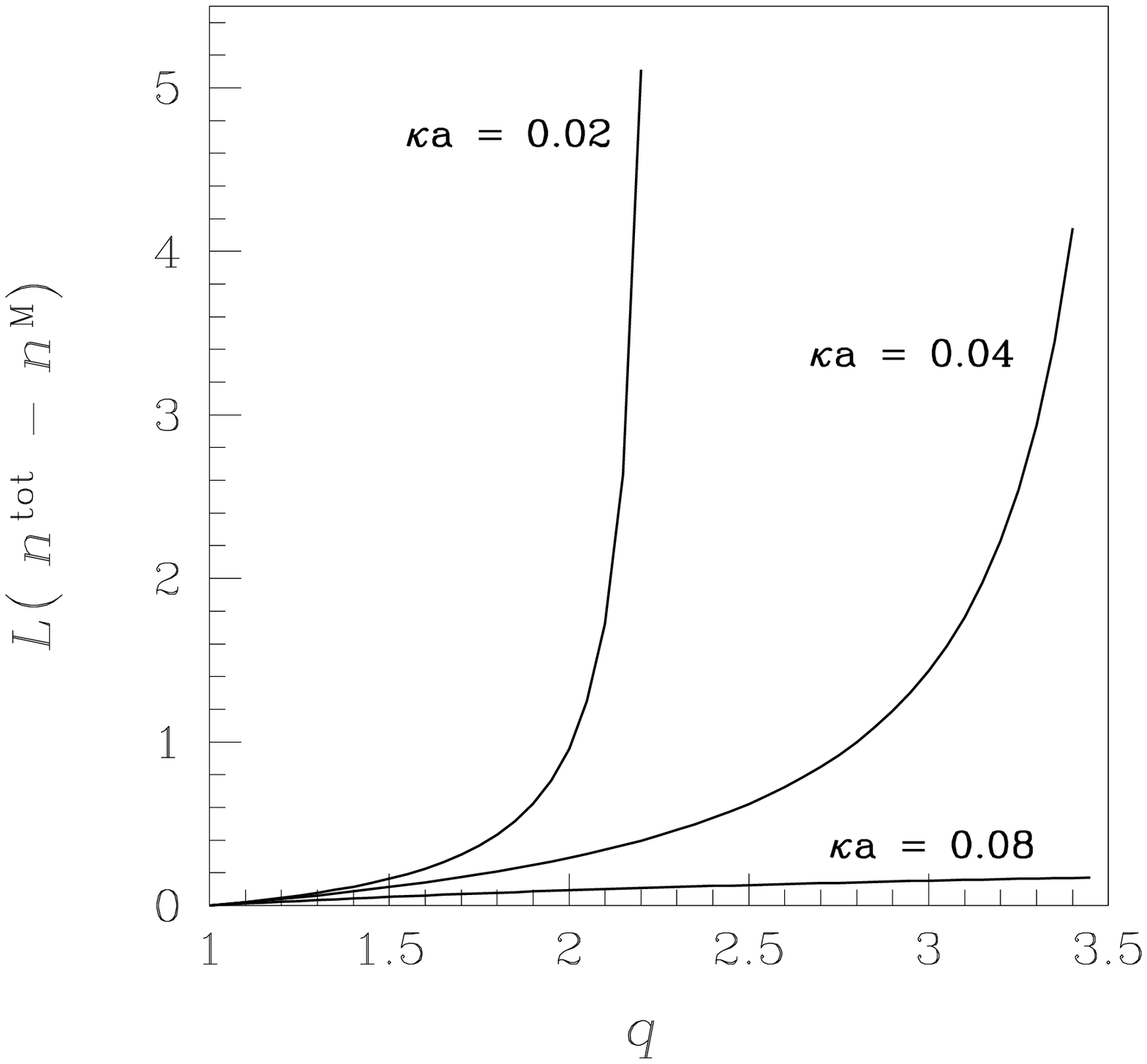} } }
\end{figure}
\centerline{\large Fig.~2}


\begin{thebibliography}{9}
\bibitem{Man_note}
Since the Manning condensation model is strictly a mean-field approximation,
it cannot be valid close to the condensation value $q= 1$ where
fluctuations dominate.
\end{thebibliography}
\end{document}